# New Cataclysmic Variable 1RXS J015017.0+375614 in Andromeda

A. Lazareva[1], N. Voroshilov[1], D. Denisenko[1,2,*], A. Kuznetsov[2], E. Gorbovskoy[2], V. Lipunov[2]

[1] International Research School, Moscow, Russia;
[2] Sternberg Astronomical Institute at Lomonosov Moscow State University, Russia;
*e-mail: d.v.denisenko@gmail.com

We report the discovery of a new cataclysmic variable in MASTER database which is identical to the faint ROSAT X-ray source 1RXS J015017.0+375614. The object was observed in outbursts to 14.9$^m$ on 2012 Nov. 28 and to 14.3$^m$ on 2013 Jan. 07, but not detected in course of the routine real-time search. Analysis of the archival MASTER data and CRTS light curve shows the large-amplitude variability from 19.1$^m$ at quiescence to 15.4-14.8$^m$ in outbursts. The new variable is most likely a dwarf nova of SU UMa subtype with rather frequent normal outbursts and occasional superoutbursts.

As of July 2013, MASTER global network of robotic telescopes (Lipunov et al., 2010) has discovered more than 350 optical transients, including ~170 new cataclysmic variables. The search procedure includes human-in-the-loop participation at the final stage with the finders checking candidate optical transients (OTs) selected automatically by the software. The criteria for the object becoming an OT candidate have been changed several times over the years of network operation, with several parameters being adjusted. For example, the minimum delta magnitude relative to the reference image has been changed from 2 to 2.5 and again to 2.2. Since MASTER survey is carried out in the white light (unfiltered), the object brightness needs to be compared to the USNO-B1.0 blue and red magnitudes using some calibration formula. It used to be $W=1.1*R2-0.1*B2$ in the beginning of survey, but after the calibration it was found that the best interpolation formula is $W=0.8*R2+0.2*B2$. This is why some objects which would have been detected as transients under the new rules did not satisfy the criteria under the old rules. Finally, some real transients were missed in the past due to the lack of manpower while checking the OT candidates in the database, while others could not have been checked because of slow internet connection to the observatories located in Blagoveshchensk, Tunka and Kislovodsk.

It was noted by one of the authors (D.D.) that at least two bright transients were actually observed by MASTER telescopes in outbursts before the CRTS or ASAS-SN detection, but were not found (and not published in time) due to the reasons above. Moreover, both such objects were located in the vicinity of previously unidentified faint ROSAT X-ray sources. Since many other cataclysmic variables were found within ROSAT error circles in the USNO-B1.0 data (Denisenko, Sokolovsky, 2011), we decided to check all MASTER-Amur candidates for the correlations with the sources from ROSAT faint catalogue (Voges et al., 2000).

The file with the coordinates and detection dates of MASTER-Amur OT candidates was prepared by A.K. The software for cross-correlating two lists of objects was written by A.L. and D.D. Since the number of objects in both lists was very large (~650 thousand OT candidates and ~105 thousand faint ROSAT sources), the index file for 1RXS catalog was created with the number of first objects for each 10x10 degree zone in R.A. and declination. Using this index file has greatly reduced the computation time: the correlations between two lists were found in less than 2 hours on the PC with the 1.6GHz dual-core processor.

The matches were searched for in the 30" circles around the faint ROSAT sources. The resulting 1200 candidates were then checked for their reality on the images in MASTER-Amur database by A.L. and N.V. under the supervision of D.D. The majority of candidates turned out to be either noise or double stars incorrectly identified in USNO-B catalogue or by MASTER pipeline. 15 candidates were already known cataclysmic variable stars or blazars. And finally, one object found by Anastasia Lazareva appeared to be a previously unknown variable star.

The new object has the coordinates $01^h50^m16^s.19$, $+37º56'20".5$ (J2000.0). It is located $11"$ from the X-ray source 1RXS J015017.0+375614 with the error radius of $15"$. It was detected in outburst on 2013 Jan. 07 with unfiltered magnitude $14.92^m$ and also on 2012 Nov. 28 at $14.27^m$. The detections by MASTER-Amur are listed in Table 1.

| Date, UT | Date, JD | Mag | Error |
|---|---|---|---|
| 2012-11-28.591 | 2456260.091 | 14.24 | 0.04 |
| 2012-11-28.623 | 2456260.123 | 14.27 | 0.04 |
| 2013-01-07.400 | 2456299.900 | 14.78 | 0.03 |
| 2013-01-07.431 | 2456299.931 | 14.92 | 0.03 |

The comparison of MASTER-Amur images taken on 2012 Aug. 21 (with the object fainter than $18.8^m$) and on 2013 Jan. 07 (with the variable in outburst at $14.8^m$) is shown in Fig. 1.

The new variable is identical to the star USNO-B1.0 1279-0035859 with the coordinates $01^h50^m16^s.183$, $+37º56'19".14$ (J2000.0) and the following magnitudes: $B1=19.00$, $R1=18.33$, $B2=19.09$, $R2=17.24$, $I=16.89$. Color-combined (BRIR) DSS finder chart is presented in Fig. 2 (10'x10' field of view). The star is listed in USNO-A2 catalogue as USNO-A2.0 1275-01092522 ($01^h50^m16^s.20$, $+37º56'19".1$, $R=18.1$, $B=18.5$). There is also an ultraviolet counterpart in GALEX database GALEX J015016.1+375619 ($FUV=18.58\pm0.08$, $NUV=18.12\pm0.05$) and the infrared detection 2MASS 01501618+3756189 ($J=15.38\pm0.05$, $H=15.18\pm0.09$, $K=14.79\pm0.09$).

The X-ray source 1RXS J015017.0+375614 has a ROSAT flux $0.0179\pm0.0096$ counts/s and hardness ratios $HR1=1.00\pm0.67$, $HR2=0.10\pm0.50$. These values are typical for the dwarf novae of $\sim18^{th}$ magnitude at quiescence and $\sim14^m$ in outbursts.

We have also checked the available data on this object in Catalina Sky Survey database (Drake et al., 2009). The object was observed 250 times from 2005 Sep. 30 to 2012 Dec. 10. Three bright outbursts were detected on 2005 Dec. 02 ($15.4^m$), 2007 Nov. 18 ($15.3^m$) and 2010 Oct. 31 ($14.8^m$). The quiescent magnitude is $19.1^m$, and there are a number of small amplitude outbursts to $16-17^m$. Overall behavior is similar to that of SU UMa-type dwarf novae with quite frequent normal outbursts which occasionally evolve into superoutbursts. The final decision on the classification of this object can be done after the analysis of its light curve during the outburst which should reveal short-period orbital variability (and possibly superhumps).

**Acknowledgments:** This object was discovered during the Astronomy project at the Sixth International Research School (IRS-6) http://www.irschool.ru held in Moscow region from 2013 June 23rd to July 3rd. The School is supported by Moscow City Center of Children and Youth Creativity. MASTER project http://observ.pereplet.ru is partially supported by State contract No. 11.G34.31.0076 and State Contract No. 14.518.11.7064 with Russian Ministry of Science and Education.

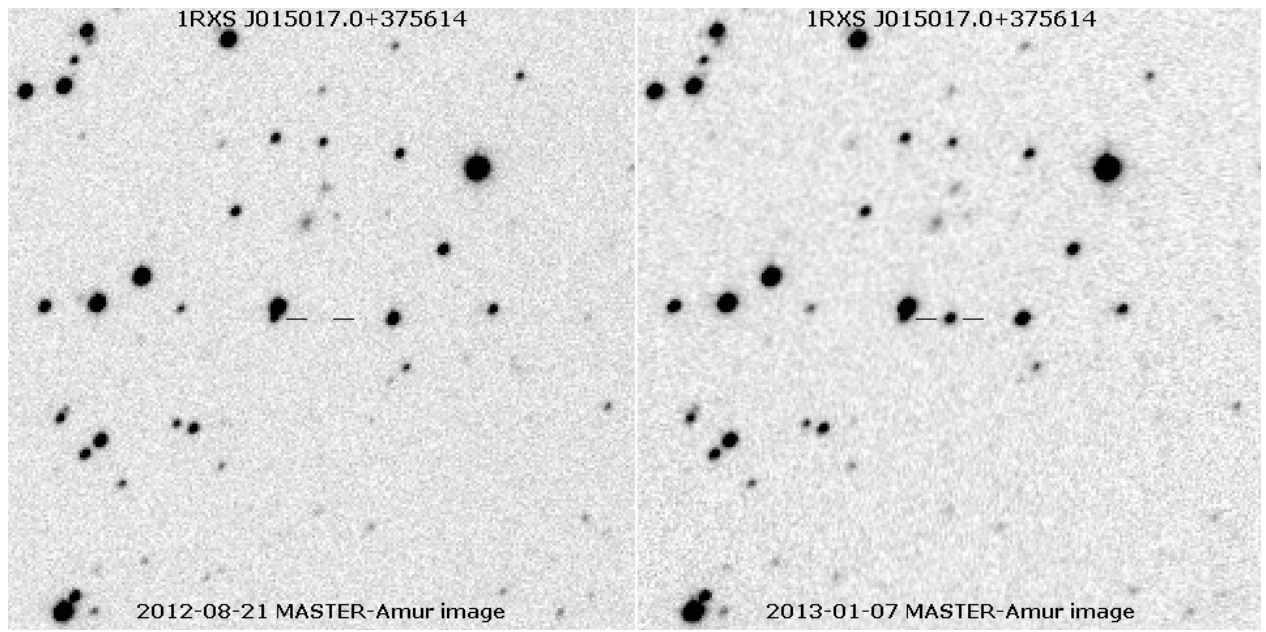

**Figure 1.** MASTER-Amur images of 1RXS J015017.0+375614 on 2012 Aug. 21 (at quiescence) and on 2013 Jan. 07 (in outburst)

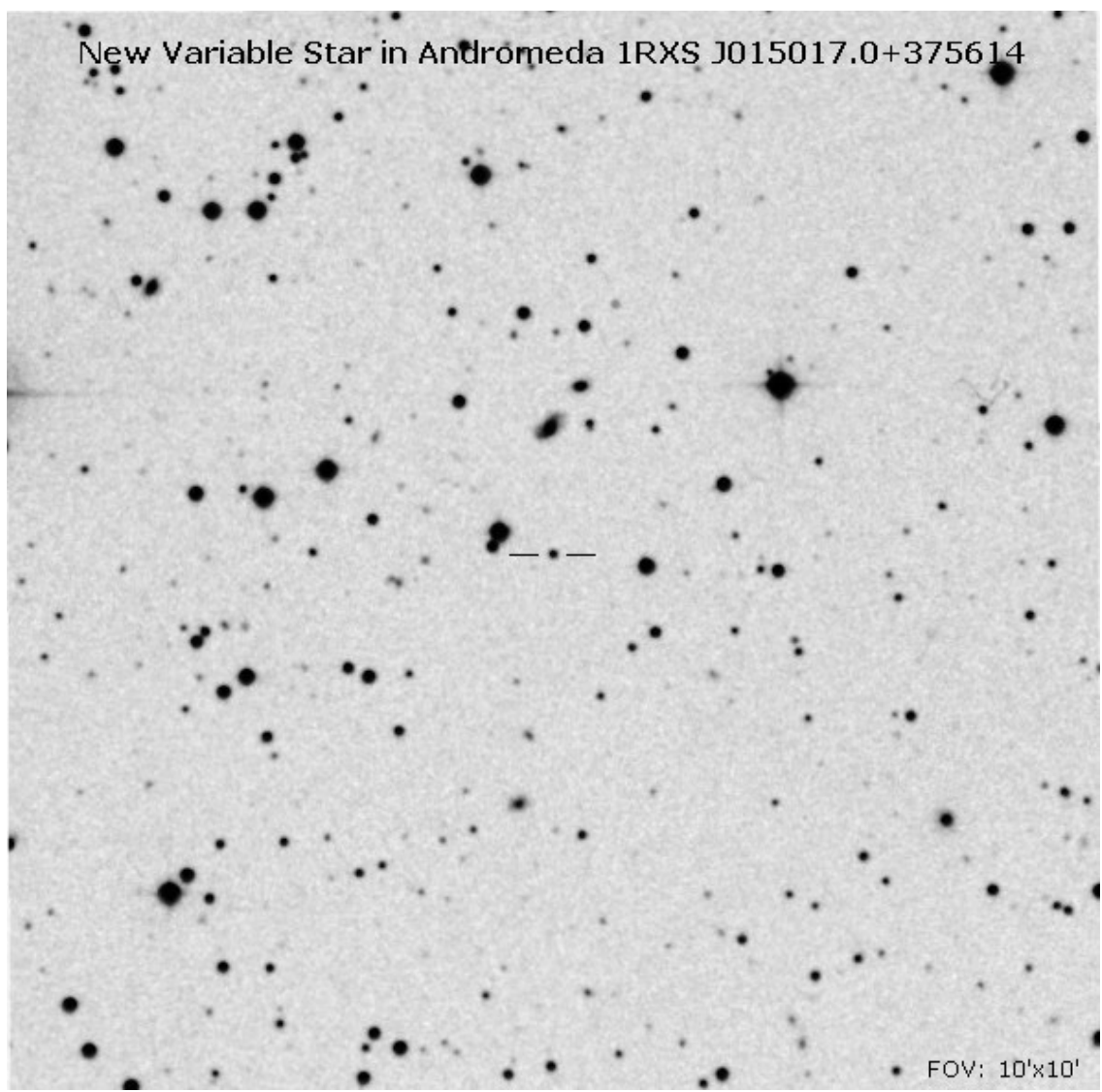

**Figure 2.** Combined DSS image of 1RXS J015017.0+375614 (10'x10' field of view)